# The structure of Tick-borne Encephalitis virus determined at X-Ray Free-Electron Lasers. Simulations

Dameli Assalauova[a] and Ivan A. Vartanyants[a]

[a] Deutsches Elektronen-Synchrotron DESY, Notkestr. 85, 22607 Hamburg, Germany

Correspondence email: dameli.assalauova@desy.de

The study of the structure of viruses by X-ray free-electron lasers (XFEL) attracts more attention in recent decades. Such experiments are based on the collection of two-dimensional diffraction patterns measured at the detector after diffraction of femtosecond X-ray pulses on biological samples. In order to prepare the experiment at the European XFEL we simulated the diffraction data for the tick-borne encephalitis virus (TBEV) with different parameters and identified their optimal values. Following necessary steps of a well-established data processing pipeline, the structure of TBEV was obtained and the efficiency of the used methods was demonstrated.

## 1. Introduction

Understanding the structure and functionality of viruses has been an important task of biology, physics, chemistry, and medicine over the last centuries. Present pandemic of COVID-19 has showed that without the knowledge of the structure and functionality of viruses it is difficult or even impossible to struggle with the SARS-Cov-2 virus. The development of different imaging techniques to study single particles, especially viruses, can bring insights into the structural features of these objects. Nowadays, several methods exist and may be applied. X-ray crystallography is the predominant method for determining the structure of biomolecules with high resolution. But since it is necessary to crystallize a protein or virus, its application is not always possible. More often this method is used for single viral proteins or compact, homogeneous, symmetric viral particles (Zhang et al., 2017; Rossmann, 2014; Yang & Rao, 2021). Small viral proteins, especially unstructured ones, are investigated by nuclear magnetic resonance (NMR) (Neira, 2013). Some membrane proteins are difficult to crystallize due to their hydrophobicity and the laborious process of manufacturing the quantities of proteins required for crystallization (Choy et al., 2021). However, knowledge of the structure of these proteins is essential for understanding the functioning of viruses. One way to solve problems mentioned above is the method of studying the spatial structure of biological particles - Single

Particle Imaging (SPI) - using cryogenic electron microscopy (Cryo-EM) (Choy et al., 2021; Egelman, 2016; Callaway, 2020). Cryo-EM makes it possible to reconstruct the structure of biological particles with resolution down to atomic or even interatomic distances. The reconstructed structure is based on many measured projections from different virions. In most cases, this method works well with structurally homogeneous particles (Agard *et al.*, 2014). However, a frequent case is when DNA/RNA is randomly packed in different virions. This leads to the blurry electron density in the center of the particle so in this case the internal structure is rather difficult to determine (Lyumkis, 2019). In addition, when using Cryo-EM, the samples must be cooled to liquid nitrogen temperature, which makes it difficult to understand the functionality of viruses at room temperatures and native environment. These limitations can be circumvented using another approach based on the use of X-ray free-electron laser (XFEL), which is in its early stages of development.

It has been shown that sufficiently intense femtosecond XFEL pulses allow obtaining diffraction patterns before radiation damage due to Coulomb explosion takes place (so-called "diffraction-before-destruction" approach (Chapman et al., 2006)). This gave rise to SPI XFEL experiments (Neutze et al., 2000; Chapman et al., 2006; Gaffney & Chapman, 2007), which were implemented at XFEL facilities built over the last 10 years. Currently, these facilities are the most powerful X-ray sources and can produce strong X-ray radiation with the pulse length of several tens of femtoseconds (Emma et al., 2010; Ishikawa et al., 2012; Decking et al., 2020; Kang et al., 2017). We should note that these sources have a high degree of coherence (Vartanyants et al., 2011; Singer et al., 2012; Gutt et al., 2012). It is this important property that makes it possible to use methods of coherent X-ray imaging (CDI) (Miao et al., 1999, 2015) in order to obtain the spatial structure of biological particles.

In spite of strong push by the community and a whole series of experiments with various biological particles (Seibert et al., 2011; Ekeberg et al., 2015; Rose et al., 2018; Kurta et al., 2017), progress in SPI experiments has not been as rapid as expected. The resolution in these experiments was limited to a few tens of nanometers. To better understand and overcome the experimental limitations, the SPI consortium was created several years ago (Aquila et al., 2015). As a result of the collaboration between the consortium members, a resolution better than 10 nm was obtained for viruses of sizes about 70 nm (Rose et al., 2018; Assalauova et al., 2020). In recent SPI experiments at the European XFEL (Hamburg, Germany) with gold nanoparticles, a resolution of 2 nm was achieved (Ayyer et al., 2021). Thus, the resolution in future SPI experiments with viruses of similar sizes can be expected to be between 2 nm and 10 nm.

During this decade, significant progress has been made in pipeline data analysis of SPI experiments to determine the electron density of viruses from the collected diffraction patterns (Gaffney & Chapman, 2007; Assalauova et al., 2020). An important step in data analysis – the clustering of two-dimensional (2D) diffraction patterns and selection of diffraction patterns that correspond to the particle under study – can be addressed with different approaches. For example, in (Assalauova et al., 2020) a clustering algorithm based on the maximum-likelihood method, which is actively used in Cryo-EM (Dempster et al., 1977), was used. In (Assalauova et al., 2022), machine learning methods using a convolutional neural network (CNN) were used for the same purpose. The reconstructed electron density of the virus was obtained by decomposing into modes of several reconstructions of the virus under study (Assalauova et al., 2020).

For a successful outcome, each SPI experiment needs careful planning. In this work, we discuss the SPI experiment planned at European XFEL with the tick-borne encephalitis virus (TBEV). Various experimental conditions will be discussed in details: incident photon flux incoming on the sample, sample-detector distance, and other parameters. The paper also presents a general data analysis pipeline and the use of the existing structure reconstruction platform (Bobkov et al., 2020). In the end, we will discuss various approaches to determine electron density on the virus from simulated data.

## 2. Tick-borne Encephalitis virus

Tick-borne encephalitis is a viral infectious disease transmitted through tick bites. The endemic area extends from west to east from the Rhine to the Urals and from north to south from Scandinavia to Italy and Greece. Tick-borne encephalitis is usually asymptomatic, but can also cause serious complications, mainly in the form of the nervous system damage. The disease can result in disability or even death. There is no cure for tick-borne encephalitis, the main preventive measure is vaccination.

The pathogen of tick-borne encephalitis is a virus belonging to the family Flaviviridae, genus Flavivirus. In addition to tick-borne encephalitis, flaviviruses cause a number of serious human diseases, including long-known infections – yellow fever, Dengue fever, West Nile fever, Japanese encephalitis, as well as newly discovered and capable of rapid spread to new territories, such as Zika fever (Barrows et al., 2018). Several million cases of flavivirus infections are reported worldwide each year (Barrows et al., 2018; Pierson & Diamond, 2020).

All viruses of this family are enveloped viruses with a virion diameter of ~50 nm. The virion core consists of a single-stranded (+)RNA molecule surrounded by protein C. It is covered on top by a lipid membrane, in which two proteins are embedded: the membrane protein M and the virion surface protein E, but protein M does not form the outer surface of the virion. Glycoprotein E is mainly responsible for the first stages of viral infection and is the target of most neutralizing antibodies (Pierson & Diamond, 2020). The structure of these viruses accounts for their natural heterogeneity: mature, immature, semi-mature and so-called "broken", i.e. deformed particles that are formed in the samples during maturation (Füzik et al., 2018; Pichkur et al., 2020). This makes it difficult to obtain the structures of flavivirus virions by X-ray crystallography, since heterogeneity prevents obtaining ordered crystals. In this regard, the method for obtaining flavivirus virion structures is Cryo-EM. Currently, the Protein Data Bank (PDB) (https://www.rcsb.org/) contains more than 40 structures of various flaviviruses.

The Cryo-EM method requires careful sample preparation (Pichkur et al., 2020) and the maintenance of a fairly high concentration of homogeneous particles of the same type, usually mature or immature virions, which are the most symmetrical. Viral particles with antigen-binding fragments (Fab-fragments) that neutralize antibodies (Long et al., 2019; Rey et al., 2018) are also studied. Two TBEV structures were obtained by the Cryo-EM method with 3.9 Å resolution (Füzik et al., 2018): the structure of the mature virion complex (structure code in PDB is 5O6A, https://www.rcsb.org/structure/5o6a) [see Fig. 1(a)] and the structure of the mature virion complex with the Fab-fragment of the mouse monoclonal antibody 19/1786 [structure code in PDB is 5O6V, https://www.rcsb.org/structure/5o6v, see Fig. 1(b)].

To model the diffraction data, we used both structures of the virus.

### 3. Data simulations for the SPI experiment on the European XFEL with TBEV

One of the goals of data simulation for TVEB was to plan the SPI experiment at the European XFEL at the Single Particles, Clusters, and Biomolecules (SPB) beamline. A typical experimental set-up of the SPI experiment is well known and described in detail in (Gaffney & Chapman, 2007). In such experiment, femtosecond X-ray pulses are scattered on single virus specimen in random orientations (see Fig. 2). The particles are destroyed after interaction with the X-ray beam, but 2D patterns containing diffraction of the initial virus state (Gorobtsov *et* *al.*, 2015) are recorded on the detector. Reconstructing the spatial structure of a virus on the basis of diffraction 2D pictures is the main task of SPI analysis.

The success of complex experiments such as SPI depends on many parameters. Parameters that can be evaluated in advance by simulation include the incident photon flux and the sample-detector distance. The scattered signal, clearly, depends on the intensity of the incident X-ray beam. At the European XFEL its intensity is 1-4 mJ per pulse, which corresponds to $10^{11}$-$10^{12}$ photons per pulse. Since in the planned experiment the size of the focal spot of the X-ray beam is 300 nm, it is natural to assume that there will be no more than $10^{12}$ photons per pulse in the focal spot.

In this experiment the planned photon energy is 6 keV (wavelength 2.07 Å). On the one hand, this energy is the lowest possible energy at the SPB station. On the other hand, the lower the energy of the incident photons, the stronger the scattered radiation. The European XFEL SPB beamline uses an AGIPD 1 Mpx detector (Allahgholi *et al.*, 2019) with a size of 1024 × 1024 pixels (one pixel size is 200 × 200 μm$^2$). The parameters of the SPB beamline described above were used in the simulation of diffraction patterns using the *MOLTRANS* program developed at DESY.

First of all, it is of interest to compare the diffraction patterns of the two available virus types: 5O6A and 5O6V. The global symmetry of both is icosahedral, but the 5O6A structure has a pronounced spherical shape, and the diffraction pattern from such an object consists of concentric rings. The structure of 5O6V looks different than that of 5O6A in reciprocal space. Due to the antigen-binding fragments, sixth-order symmetry can be seen in the diffraction patterns, indicating the appearance of characteristic features of the structure. Examples of diffraction patterns from two structures in random orientations are shown in Fig. 3.

To demonstrate other parameters, we used the 5O6V structure [see Fig. 1(b)]. The scattered signal recorded by the detector depends on the parameters of the experimental setup as well as on the sample – the larger the object, the higher the intensity of the scattered signal. Note that the Cryo-EM structures taken from the PDB bank and used in the present work describe the surface protein E and membrane protein M and do not characterize the inner nucleocapsid formed by protein C and the RNA structure chain. Naturally, in the SPI experiment, the presence of the RNA nucleocapsid in the particles will contribute to the scattered signal on the detector.

Fig. 4 shows three diffraction patterns from a single virus with different photon flux in the focus of the X-ray beam. The figure presents that the diffraction pattern contains more structural features of the particle with increasing of the photon flux from $10^{10}$ to $10^{12}$ photons at the X-ray beam focus. If the scattering signal from the virus is too weak [see Fig. 4(a)], the features of the object are less visible, making further data analysis steps difficult. The minimum number of recorded photons per diffraction pattern needed for a successful reconstruction was actively studied (Ayyer *et al.*, 2015, 2019; Giewekemeyer *et al.*, 2019; Ekeberg *et al.*, 2022). It was shown (Poudyal *et al.*, 2020) that low intensity of diffraction pattern can be compensated with the number of the patterns collected during the experiment. With the parameters used to obtain diffraction pattern in Fig 4(c), the scattered signal is determined to a value of 1.24 nm$^{-1}$ in reciprocal space, which corresponds to a resolution of 5 nm in real space. From the simulations (see Fig. 4) we can conclude that the maximum signal intensity of $10^{11}$-$10^{12}$ photons at the X-ray beam focus, which is achievable at the European XFEL, is necessary for the experiment to succeed.

Another important parameter of the SPI experiment that can be analysed with the simulations is the sample-detector distance. If the distance is too small, it will not allow to obtain a pronounced diffraction from the sample, but it will allow to obtain a high resolution. If the distance is too large, it will not allow to achieve the desired resolution. Examples of diffraction patterns with different distances from 1 m to 3 m are shown in Fig. 5(a)-5(c). An angle-averaged intensity was plotted for each case and is shown in Fig. 5(g)-5(i). As it was expected, at the shortest distance of 1 m, the diffraction pattern shows the characteristic features of the virus structure, and the resolution in the real space reaches 2 nm. As the distance increases to 2 m, these features become more pronounced, but the resolution in real space drops to 4 nm. At the maximum distance of 3 m, the diffraction pattern from the virus is clearly distinguishable; in Fig. 5(i) the characteristic rings are clearly visible. But the resolution in reciprocal space is limited to 6.3 nm.

An important factor when choosing the optimal sample-detector distance is the structure of the detector panels. Their configuration highly depends on the sample and experimental set-up. In practice, the detector panels are not placed together; there is a distance between them. There is also a gap in the center of the detector for the direct (central) beam to pass through. Experimental diffraction patterns with the superimposed geometry of gaps between the detector panels are shown in Fig. 5(g)-5(e). The detector geometry was taken from one of the SPI experiments at the SPB beamline of European XFEL. The figure shows that a large part of the central diffraction peak at a distance of 1 m is not determined because of the panel positions in the central part of the detector. Information about the size of the central peak must be recorded and is essential when reconstructing the object. It is also important to take this into account when planning the experiment, in particular, when choosing the optimal distance. From the analysis of the performed simulations (see Fig. 5) with different sample-detector distances, we can conclude that a distance of 2-3 m is preferable. In this case, the detector geometry makes it possible to distinguish all structural features of the virus, and the momentum transfer vector *q* reaches a value of 1.04-1.55 $nm^{-1}$ in reciprocal space, which corresponds to a resolution of 4-6 nm in real space.

## 4. Data analysis pipeline of SPI experiment to obtain a spatial structure of the object

The data analysis pipeline of SPI experiments is aimed at obtaining the three-dimensional (3D) structure of the object on the basis of diffraction patterns collected in the experiment. The structure of the object is determined by the electron density distribution, and measured 2D diffraction patterns contain information about the reciprocal space, which is a 3D Fourier transform image of the electron density. By determining the orientation of the series of 2D diffraction patterns, they can be formed into one 3D volume in reciprocal space. Then the scattering phase values need to be recovered, which allows obtaining the electron density distribution through the Inverse Fourier transform.

To account for the specifics of the experiment, additional steps were included in the data analysis procedure (Rose *et al.*, 2018; Bobkov *et al.*, 2020; Assalauova *et al.*, 2020). First, due to the fact that only a small fraction of the images contains diffraction patterns, empty patterns are filtered out from the analysis. Second, before starting the analysis of the diffraction patterns, the position of the center of the diffraction pattern relative to the detector position should be precisely determined. Then, among all diffraction patterns, the ones which contain scattering signal from a single virus are selected. For this purpose, first the size of the samples is estimated from the diffraction patterns and those with the reasonable size distribution are selected. Then, the selected patterns are classified according to the features of the diffraction patterns, which are related to the features of the sample structure in real space. Thus, the patterns related only to the studied object are distinguished – so-called single hits. From our experience, significant part of diffraction patterns refers to impurities or water droplets, a part of patterns contains diffraction from several objects combined. Quality of diffraction patterns classification directly influences the obtained structure resolution – if the quality of single hits classification is insufficient, structure reconstruction becomes impossible.

Selected diffraction patterns of the studied object are combined in reciprocal 3D space. For this purpose, the orientations of the diffraction patterns relative to each other are determined. They are respectively related to the orientations of the object inserted to the X-ray pulse during the experiment. The orientation determination and the reconstruction of the reciprocal space volume are based on the maximum-likelihood method (Loh & Elser, 2009) implemented into Expand-Maximize-Compress (EMC) algorithm (Ayyer *et al.*, 2016). It proved to work well and to orient relatively small amount of diffraction data with missing areas and small number of photons. After the reconstruction of the reciprocal space volume, the background signal is corrected. It is usually caused by parasitic scattering on the elements of the experimental setup. The background signal does not depend on the orientation of the sample, so the background correction is performed at 3D intensity volume in reciprocal space. At this stage, influence of the background is averaged, and it is easier to correct it.

Next, the phases of the diffraction patterns are reconstructed in reciprocal space and the structure is reconstructed in real space. For this purpose, iterative phase retrieval algorithms are used (Fienup, 1982, 2013; Marchesini *et al.*, 2003). Finally, the resolution of the resulting electron density distribution is evaluated.

The presented pipeline was tested with the data of several SPI experiments on XFELs: LCLS (Stanford, USA) and European XFEL (Sobolev *et al.*, 2020). Processing the experimental SPI data according to the presented scenario allowed to obtain a significant improvement in the structure recovery in SPI (Rose *et al.*, 2018; Assalauova *et al.*, 2020). Software was developed to implement all the stages of the pipeline, and a platform for automated data processing of SPI XFEL experiments was created (Bobkov *et al.*, 2020). The platform includes containerized software that is integrated into a software pipeline, which provides automatic processing from experimental data to structure reconstruction. The platform can be quickly run on any computing architecture with container support (e.g., Docker, (Merkel, 2014)), as well as in computing clusters running by Kubernetes (https://kubernetes.io/). Information about the developed software (Bobkov *et al.*, 2020) and the automatic data processing platform is publicly available (https://gitlab.com/spi_xfel).

## 5. Spatial structure of the TBEV from simulated data

In this work, we adapted the mentioned above data analysis pipeline for simulated diffraction patterns from TBEV with and without Fab-fragments. The goal of SPI experiment is from diffraction patterns collected during the experiment to obtain the spatial structure of the TBEV. Considering analysis performed in the Section 3, a 1000 diffraction patterns (without superimposed detector mask as shown in Fig 5 (a)-(c)) were created for each TBEV type (see Fig. 1). The detector parameters: detector size 128 × 128 pixels, pixel size 1.6 × 1.6 mm$^2$ were used in the simulations and represented binned configuration. Such dimensions were set in order to match the real size of the AGIPD 1 Mpx detector but also to save computational time. Other parameters: the sample-detector distance 2.1 m, signal intensity 10$^{11}$ photons in focus in random orientations. The distance was chosen according to the geometry of the planned experiment at the SPB beamline of European XFEL. The number of simulated diffraction patterns (1000) was based on the existing research (Poudyal *et al.*, 2020) showing the dependence between this number, experimental parameters, and spatial resolution of several nanometers or less. The number of simulated diffraction patterns can be also estimated from the experience with experimental data (Rose *et al.*, 2018; Assalauova *et al.*, 2020, 2022). This number depends on the studied particle and experimental conditions.

Since only the structure of the TBEV obtained by Cryo-EM (Füzik *et al.*, 2018) was used in the simulation, clustering and classification of the diffraction patterns by object type was not required. Since the orientation of the particle in the simulations is known, combining the data into a diffraction 3D intensity volume in reciprocal space was done according to the identified orientations of the virus. The result of the simulations in reciprocal space is shown in Fig. 6. For virus 5O6A [see Fig. 6(a)], as expected, concentric rings are observed. Due to the presence of Fab-fragments in the structure of virus 5O6V, diffraction fringes in reciprocal space are observed [see Fig. 6(b)].

The experimental background signal was not additionally included to the diffraction patterns, as it strongly depends on the experiment and the set-up. The next step is to reconstruct the scattering phases and structure of the object. As described earlier, iterative phase retrieval algorithms are used for this task (Fienup, 1982, 2013; Marchesini *et al.*, 2003). These algorithms are based on the Fourier transform between real and reciprocal spaces using two constraints: in reciprocal space, the signal amplitude is set equal to the experimentally measured values, and in real space, the object occupies a limited volume whose approximate size is known in advance.

To obtain the spatial structure of the virus (for 5O6A and 5O6V) the following combination of algorithms was used: 100 iterations of Continuous Hybrid-Input-Output, followed by 200 iterations of Error Reduction with an alternating Shrink Wrap algorithm every 10 iterations with a threshold value of 0.2. This combination of algorithms was repeated 4 times for one reconstruction with the total number of iterations 1200. A total of 30 reconstructions were made. Then they were averaged using the mode decomposition method described in (Assalauova *et al.*, 2020). The main mode was further considered as the final spatial structure of the TBEV, shown in Fig. 7.

The spatial structure of 5O6A and 5O6V has a ring shape with no density inside the particle, as in the data used for simulation of diffraction patterns. From the obtained result, we can see that due to the low resolution Fab-fragments in the structure of 5O6V cannot be distinguished. Thus, the size of both structures is about 60 nm. This corresponds to the size of the 5O6V structure (~57 nm) obtained with Cryo-EM (Füzik *et al.*, 2018). At the same time, the size of the virus obtained by the reconstruction is larger than the size of the 5O6A (~47 nm, (Füzik *et al.*, 2018)). Note, that the structures 5O6A and 5O6V used for simulations did not contain electron density inside (internal RNA).

## 6. Summary and outlook

An analysis of diffraction patterns simulation for the SPI experiment is presented. SPI method allows obtaining the spatial structure of biological nanoparticles using intense XFEL femtosecond pulses. Such experiments require careful preparation and planning. To prepare the experimental set-up and efficiently use the beamtime, some parameters can be estimated in advance, for example, by means of diffraction patterns simulation.

A TBEV was chosen as a studied object, for which the structure of the outer envelope was known from Cryo-EM (Füzik *et al.*, 2018). The size and relative homogeneity of TBEV make it a good object of study in SPI experiments on XFEL.

Two TBEV structures were used to simulate diffraction patterns: a mature virion complex (structure 5O6A in PDB) and a mature virion with a Fab-fragment (structure 5O6V in PDB). These two structures give different diffraction patterns in reciprocal space. In the case of 5O6A, only concentric rings were distinguishable in diffraction patterns. Whereas for 5O6V one can observe characteristic features of the structure associated with "spikes" of Fab-fragments on the surface of the viral particle.

In order to prepare the SPI experiment at the European XFEL, the following parameters were varied during diffraction patterns simulations: the x-ray beam intensity at the focus and the sample-detector distance. With the help of simulations, it was possible to determine the optimal parameters and use them in the preparation of the experiment. For the SPB beamline of European XFEL the following parameters were identified as optimal: signal intensity $10^{11}$-$10^{12}$ photons in the focus, sample-detector distance 2-3 m.

Only necessary steps of SPI data analysis pipeline were used for the simulation data: merging the data into a diffraction 3D volume; scattering phases reconstruction and object structure reconstruction. Using iterative phase retrieval algorithms, 30 virus reconstructions were obtained, they were averaged by mode decomposition, and the final structure for 5O6A and 5O6V was chosen as the main mode of decomposition. The result of the analysis was the following. For the 5O6V structure it is impossible to distinguish Fab-fragments, for both structures (5O6A and 5O6V ) a ring corresponding to the virus membrane was present with no density inside, which corresponds to the original structure used for simulations.

In this study, several steps of SPI data analysis pipeline were not used, such as background correction, single hit diffraction pattern classification, orientation determination due to the known parameters of the simulations. In the real SPI experiments, these steps cannot be avoided and has to be performed very carefully in order to obtain the final particle structure with high resolution. Each of the steps mentioned above became the study of the separate research, such as for single hit classification (Bobkov *et al.*, 2015; Shi *et al.*, 2019; Cruz-Chú *et al.*, 2021; Ignatenko *et al.*, 2021; Assalauova *et al.*, 2022), orientation determination (Loh & Elser, 2009; Ayyer *et al.*, 2016), and background subtraction (Rose *et al.*, 2018; Kurta *et al.*, 2017; Lundholm *et al.*, 2018).

The presented study using simulated data shows the potential for SPI experiments which ultimate goal is high-resolution imaging. While electron ptychography single particle imaging recently achieved a resolution of 0.2 angstroms (Chen *et al.*, 2020), SPI at XFEL still have not reached the atomic resolution. Higher the number of photons scattered on the sample in SPI can increase the diffracted signal, however, this approach has its limit and the signal cannot raise infinitely (Gorobstov *et al.*, 2015). One of the factors is the number of collected diffraction patterns during the experiment. Megahertz facilities, such as the European XFEL (Decking *et al.*, 2020; Mancuso *et al.*, 2019), allow to obtain the necessary amount of data in a shorter time. The first experiments have already demonstrated the feasibility of megahertz SPI data acquisition (Sobolev *et al.*, 2020), which plays a great role for future progress in SPI experiments performed at XFEL. Sufficient number of diffraction patterns with appropriate photon signal recorded by special detectors, effective sample delivery system, and novel data analysis techniques for single particle classification and orientation determination can open new possibilities of SPI technique with XFEls and reach its goal of high-resolution imaging of biological particles.

## 7. Acknowledgements

The authors acknowledge fruitful discussions with S.A. Bobkov, V.R. Samygina, A.B. Teslyuk, and V.A. Ilyin that helped to shape this paper. The authors also acknowledge the support provided by E. Weckert for the opportunity to use the MOLTRANS program to simulate the diffraction data. We acknowledge the careful reading of the paper by A. Shabalin.

Dammann, J., Danared, H., de Zubiaurre Wagner, A., Delfs, A., Delfs, T., Dietrich, F., Dietrich, T., Dohlus, M., Dommach, M., Donat, A., Dong, X., Doynikov, N., Dressel, M., Duda, M., Duda, P., Eckoldt, H., Ehsan, W., Eidam, J., Eints, F., Engling, C., Englisch, U., Ermakov, A., Escherich, K., Eschke, J., Saldin, E., Faesing, M., Fallou, A., Felber, M., Fenner, M., Fernandes, B., Fernández, J. M., Feuker, S., Filippakopoulos, K., Floettmann, K., Fogel, V., Fontaine, M., Francés, A., Martin, I. F., Freund, W., Freyermuth, T., Friedland, M., Fröhlich, L., Fusetti, M., Fydrych, J., Gallas, A., García, O., Garcia-Tabares, L., Geloni, G., Gerasimova, N., Gerth, C., Geßler, P., Gharibyan, V., Gloor, M., Głowinkowski, J., Goessel, A., Gołębiewski, Z., Golubeva, N., Grabowski, W., Graeff, W., Grebentsov, A., Grecki, M., Grevsmuehl, T., Gross, M., Grosse-Wortmann, U., Grünert, J., Grunewald, S., Grzegory, P., Feng, G., Guler, H., Gusev, G., Gutierrez, J. L., Hagge, L., Hamberg, M., Hanneken, R., Harms, E., Hartl, I., Hauberg, A., Hauf, S., Hauschildt, J., Hauser, J., Havlicek, J., Hedqvist, A., Heidbrook, N., Hellberg, F., Henning, D., Hensler, O., Hermann, T., Hidvégi, A., Hierholzer, M., Hintz, H., Hoffmann, F., Hoffmann, M., Hoffmann, M., Holler, Y., Hüning, M., Ignatenko, A., Ilchen, M., Iluk, A., Iversen, J., Iversen, J., Izquierdo, M., Jachmann, L., Jardon, N., Jastrow, U., Jensch, K., Jensen, J., Jeżabek, M., Jidda, M., Jin, H., Johannson, N., Jonas, R., Kaabi, W., Kaefer, D., Kammering, R., Kapitza, H., Karabekyan, S., Karstensen, S., Kasprzak, K., Katalev, V., Keese, D., Keil, B., Kholopov, M., Killenberger, M., Kitaev, B., Klimchenko, Y., Klos, R., Knebel, L., Koch, A., Koepke, M., Köhler, S., Köhler, W., Kohlstrunk, N., Konopkova, Z., Konstantinov, A., Kook, W., Koprek, W., Körfer, M., Korth, O., Kosarev, A., Kosiński, K., Kostin, D., Kot, Y., Kotarba, A., Kozak, T., Kozak, V., Kramert, R., Krasilnikov, M., Krasnov, A., Krause, B., Kravchuk, L., Krebs, O., Kretschmer, R., Kreutzkamp, J., Kröplin, O., Krzysik, K., Kube, G., Kuehn, H., Kujala, N., Kulikov, V., Kuzminych, V., La Civita, D., Lacroix, M., Lamb, T., Lancetov, A., Larsson, M., Le Pinvidic, D., Lederer, S., Lensch, T., Lenz, D., Leuschner, A., Levenhagen, F., Li, Y., Liebing, J., Lilje, L., Limberg, T., Lipka, D., List, B., Liu, J., Liu, S., Lorbeer, B., Lorkiewicz, J., Lu, H. H., Ludwig, F., Machau, K., Maciocha, W., Madec, C., Magueur, C., Maiano, C., Maksimova, I., Malcher, K., Maltezopoulos, T., Mamoshkina, E., Manschwetus, B., Marcellini, F., Marinkovic, G., Martinez, T., Martirosyan, H., Maschmann, W., Maslov, M., Matheisen, A., Mavric, U., Meißner, J., Meissner, K., Messerschmidt, M., Meyners, N., Michalski, G., Michelato, P., Mildner, N., Moe, M., Moglia, F., Mohr, C., Mohr, S., Möller, W., Mommerz, M., Monaco, L., Montiel, C., Moretti, M., Morozov, I., Morozov, P., Mross, D., Mueller, J., Müller, C., Müller, J., Müller, K., Munilla, J., Münnich, A., Muratov, V., Napoly, O., Näser, B.,

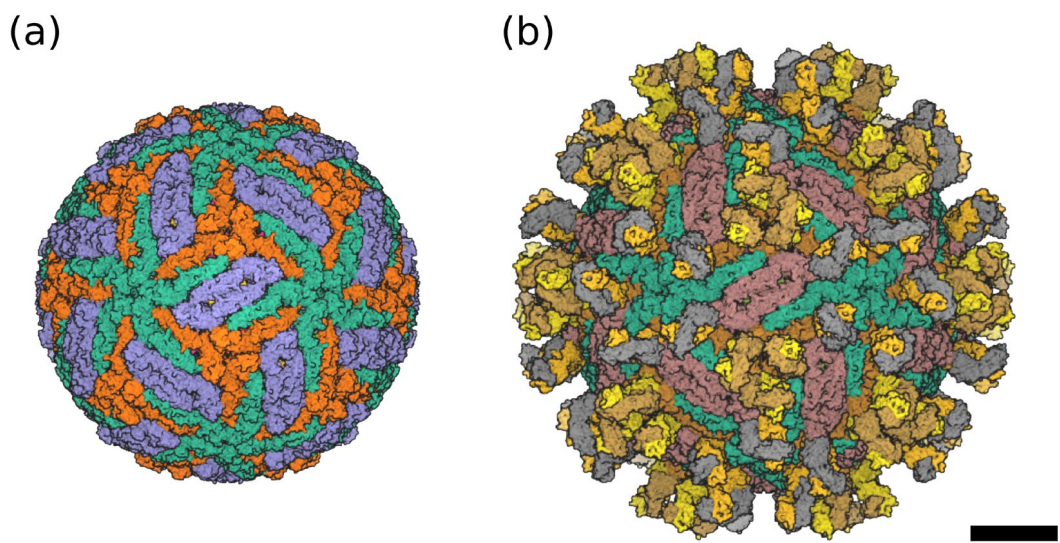

**Figure 1.** Cryo-EM structure of TBEV. (a) The structure of the mature TBEV particle 5O6A. (b) Structure of the complex with Fab-fragment of neutralizing monoclonal antibody 5O6V. The structures of TBEV were taken from the Protein Data Bank [37, 38]. The size of the scale bar is 10 nm.

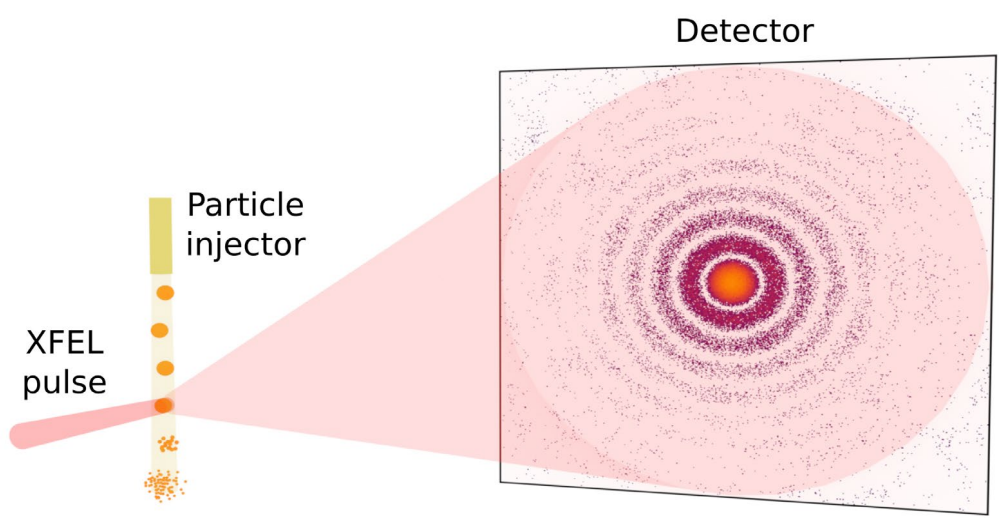

**Figure 2.** Experimental set-up of SPI-experiment to determine the spatial structure of single biological particles. The specimen of the particle is injected into an X-ray laser beam in a random orientation, and the scattered radiation is recorded by the detector.

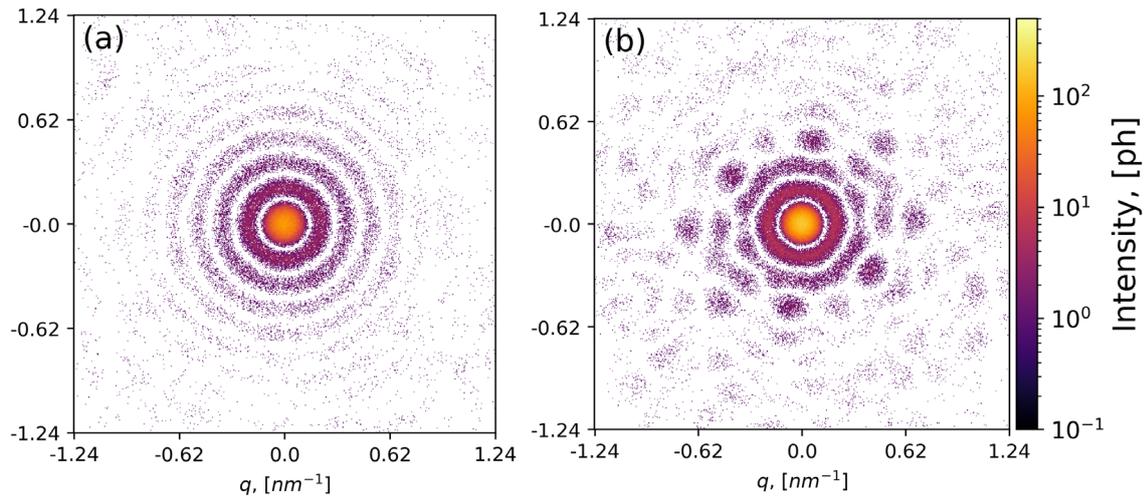

**Figure 3.** Diffraction patterns from single objects in random orientation: (a) 5O6A structure [for TBEV in Fig. 1(a)]; (b) 5O6V structure [for TBEV with Fab-fragments in Fig. 1(b)]. Simulations parameters: wavelength 2.07 Å, X-ray beam focus 300 nm, detector 512 × 512 pixels, pixel size 400 × 400 μm², distance 2.5 m, signal intensity $10^{12}$ photons in focus.

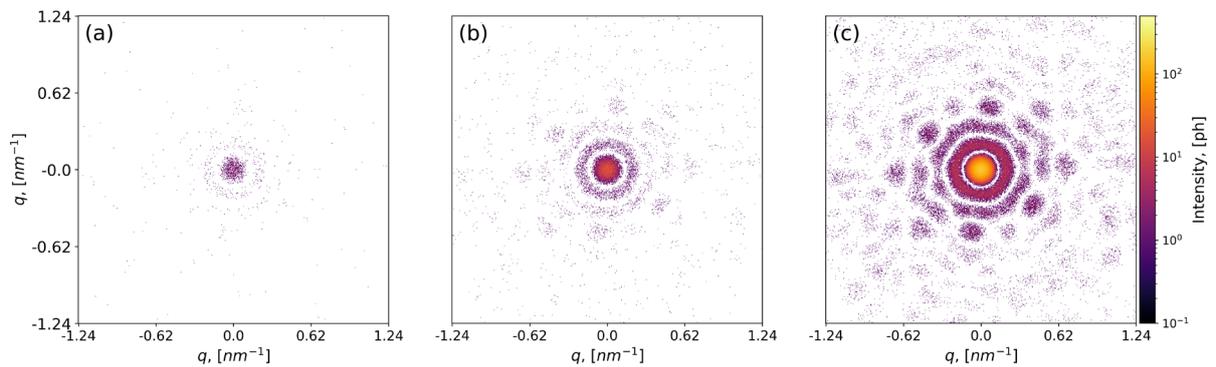

**Figure 4.** Diffraction patterns from a single TBEV in random orientation. Signal intensities: (a) $10^{10}$, (b) $10^{11}$, (c) $10^{12}$ photons in focus. The sample-detector distance is 2.5 m.

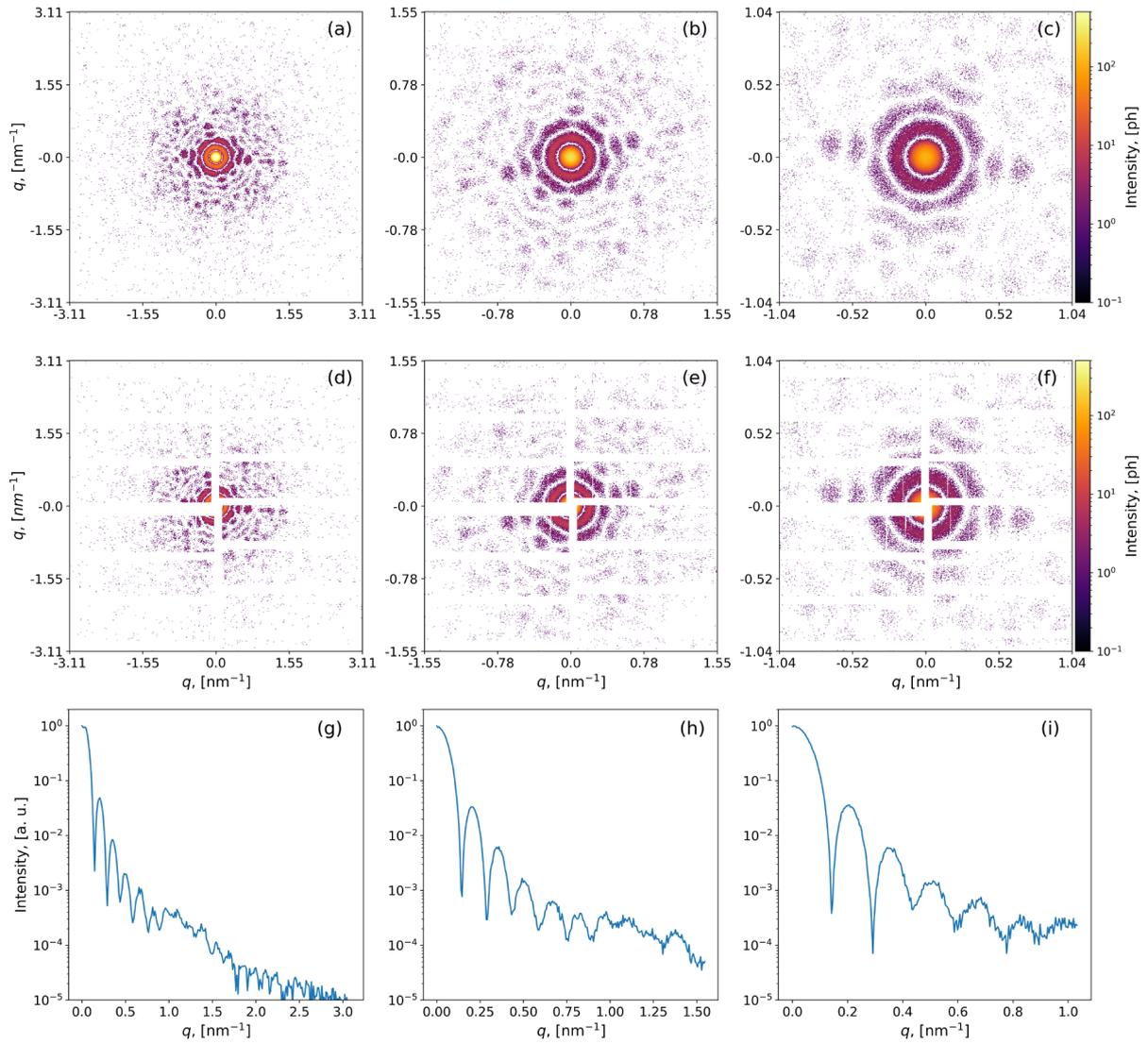

**Figure 5.** Diffraction patterns from a single TBEV in random orientation. At distances of 1 (a, d, g), 2 (b, e, h) and 3 (c, f, i) m. Examples of diffraction patterns with the detector mask superimposed on them (d)-(f). Functions (g)-(i) corresponding to the angular averaged intensity in (a)-(c).

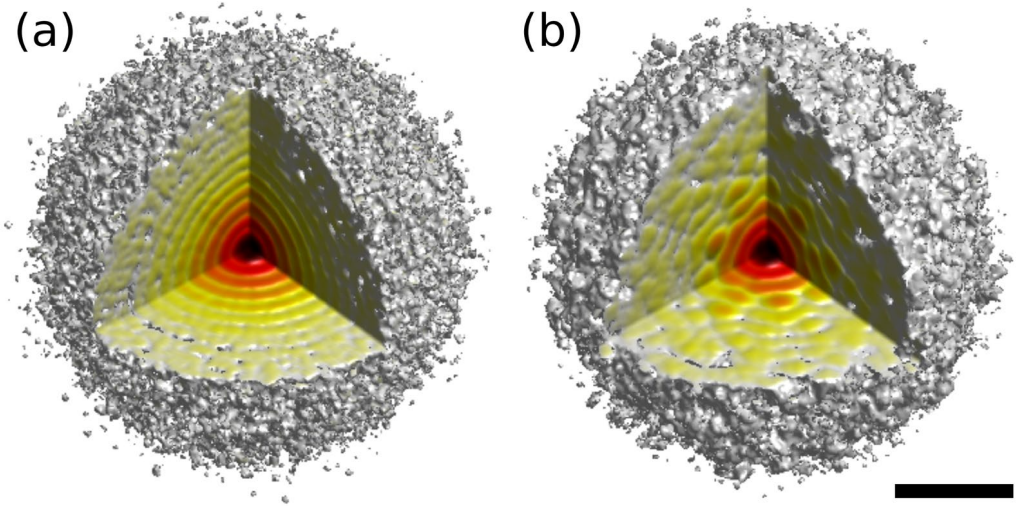

**Figure 6.** Diffraction 3D intensity volume in reciprocal space: (a) 5O6A structure [for TBEV in Fig. 1(a)]; (b) 5O6V structure [for TBEV with Fab-fragments, Fig. 1(b)]. The size of the scale bar is 1 nm-1.

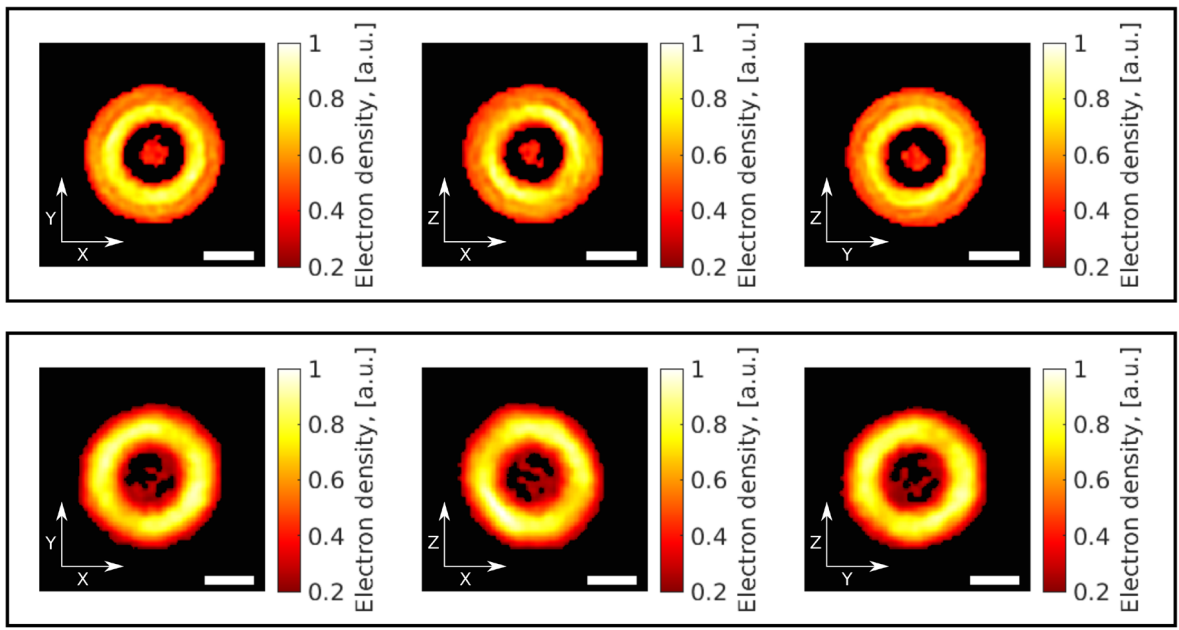

**Figure 7.** Central slices of the TBEV reconstructions. Upper row for structure 5O6A, lower row for structure 5O6V. Electron density values are normalized to the maximum, values less than 0.2 are shown in black. The size of the scale bar is 20 nm.